\documentclass{article}

    \PassOptionsToPackage{numbers, compress}{natbib}
    \usepackage[preprint]{neurips_2020}

\usepackage[utf8]{inputenc} % allow utf-8 input
\usepackage[T1]{fontenc}    % use 8-bit T1 fonts
\usepackage{hyperref}       % hyperlinks
\usepackage{url}            % simple URL typesetting
\usepackage{booktabs}       % professional-quality tables
\usepackage{amsfonts}       % blackboard math symbols
\usepackage{nicefrac}       % compact symbols for 1/2, etc.
\usepackage{microtype}      % microtypography
\usepackage[toc,page]{appendix}
\usepackage{graphicx}

\title{ML-based Flood Forecasting: Advances in Scale, Accuracy and Reach}

\author{%
  Sella Nevo \\
  Google Research\\
  \texttt{sellanevo@google.com} \\
  \And
  Gal Elidan \\
  Google Research\\
  \texttt{elidan@google.com} \\
  \And
  Avinatan Hassidim \\
  Google Research\\
  \texttt{avinatan@google.com} \\
  \And
  Guy Shalev \\
  Google Research\\
  \texttt{guysha@google.com} \\
  \And
  Oren Gilon \\
  Google Research\\
  \texttt{ogilon@google.com} \\
  \And
  Grey Nearing \\
  Google Research\\
  \texttt{gsnearing@google.com} \\
  \And
  Yossi Matias \\
  Google Research\\
  \texttt{yossi@google.com} \\
}

\begin{document}

\maketitle

\begin{abstract}
  Floods are among the most common and deadly natural disasters in the world, and flood warning systems have been shown to be effective in reducing harm. Yet the majority of the world's vulnerable population does not have access to reliable and actionable warning systems, due to core challenges in scalability, computational costs, and data availability. In this paper we present two components of flood forecasting systems which were developed over the past year, providing access to these critical systems to 75 million people who didn't have this access before.
\end{abstract}

\section{Introduction}

\subsection{Flooding harms and early warning impact}
Floods are the most common, and among the most deadly, natural disaster on the planet, affecting hundreds of millions of people every year, and causing between thousands and tens of thousands of fatalities annually \cite{cred2015human}. Recent research shows that those who receive flood warnings prior to the arrival of the flood are twice as likely to evacuate or take other protective measures than those who don't receive such warning, with 65\% of those warned taking action \cite{yale}.

As a result, warning systems have the potential to save lives and reduce harm to livestock and other assets. For this reason, the World Bank has identified early warning systems for floods as the most cost-effective tool for climate change adaptation amongst all options evaluated. It estimated that for each dollar spent on such systems, nine dollars of damages are prevented \cite{worldbank}. Machine learning is well suited to help provide reliable operational flood forecasts at large scales \cite{Nevo2019ML-FF-Scale}.

\subsection{Challenges}

The vast majority of populations vulnerable to flooding do not have access to reliable, actionable flood warnings \cite{hirpa2016saving}. There are several reasons why this is a difficult challenge, including the following:

\begin{itemize}

\item \emph{Local calibration}: Existing high-accuracy flood models often require large amounts of per-site data and effort to set up and calibrate \cite{beven2011rainfall}. This limits the scale at which such systems can be deployed, and therefore most high-resolution, high-accuracy systems are deployed at city-scale rather than country, continent, or global scales.

\item \emph{Computational Complexity}: Another challenge for scaling actionable flood forecasting systems are the computational costs required to make alerts targeted and useful. Inundation/hydraulic/hydrodynamic modeling requires computational power that scales linearly with coverage area, but is also inversely proportional to the cube of the resolution. Since the number of rivers rises exponentially with the reduction of their width \cite{downing2012global}, this effectively means that such systems require exponential growth in computational costs.

\item \emph{Data Scarcity}: Flood forecasting systems rely on a wide range of inputs that can be critical for accuracy, such as discharge, riverbed bathymetry, elevation maps, etc. While this data is publicly available in a small number countries, many countries do not collect it at all, and many of those that do refuse to make it publicly available and may even define it as classified. This makes reliable forecasting a significant challenge in many regions.

\end{itemize}

\subsection{Paper overview}

This paper focuses on riverine flooding, which is responsible for a majority of flood fatalities globally \cite{jonkman2005global}. Our long term vision is for everyone in the world to have access to reliable riverine flooding alerts. We discuss two steps in that direction that specifically help address some of the issues outlined above:

\begin{itemize}

\item \emph{Water level-based hydrologic models}: Within the last year, ML has been shown to be able to help address the local calibration problem \cite{kratzert2019towards,kratzert2019toward}. Here we explore adjusting similar models to work with water level data, which is more common globally than discharge data. This expands the globally-available data set for streamflow modeling.

%\item \emph{Pure-ML inundation models}: A new method for inundation modeling, which performs reasonably well with no elevation maps, and no human intervention from data to deployment. This simple and scalable system allowed us to launch high-accuracy inundation models across more than 250,000 square kilometers within several months of development, and can scale several orders of magnitude further.

\item \emph{Morphological models}: A new method for inundation modeling, which combines physics-based and ML modeling, requires less manual calibration, is more computationally efficient, and achieves better accuracy in real-world data-scarce conditions than classic hydrodynamic modeling using finite-element solutions. 

\end{itemize}

These models were launched in India and Bangladesh covering an area of over 67,000 square kilometers protecting 75 million people (though some areas only have access to one of the models). In the past 5 months, we have sent over 39 million alerts to people affected by the floods and governmental relief agencies.

%In this paper, we share quantitative metrics and results for all three of these conceptually new models. This includes benchmarking against existing classical models, as well as real-world metrics from our operational systems.

\section{Water level-based hydrologic models}
\label{section::stage}
Hydrologic models aim to forecast the amount of water in a river by taking inputs such as precipitation, radiation, and other meteorological variables to determine discharge, which is a volumetric rate, at a timescale of hours, days, or months. Most literature about hydrologic modeling assumes discharge as the variable of interest, with a few exceptions \cite{moshe2020hydronets}. Discharge measurements are typically derived by measuring the height of water in a river (called \textit{stage}) and then using a \textit{rating curve} to translate this measurement into an estimate of discharge. Rating curves are derived from bathymetry and velocity measurements taken in a particular river, and must be routinely updated as river conditions change. Most rivers around the world do not have rating curves, and calibrating models to water level (stage) is fundamentally difficult because the relationship between precipitation and water level is strongly controlled by local channel geometry. This means that the majority of the world's streamflow data is unavailable for calibrating hydrologic models, which presents a challenge for designing operational flood warning systems that scale globally.

\subsection{Deployed model}
Machine learning can help address this problem. Currently, we have deployed water level models in operational flood warning systems in India and Bangladesh covering 67,000 square kilometers and 75 million people, where reliable discharge measurements are not available. These models are based on regressions that use recent, near real-time water level measurements from upstream gauges as inputs, going back 72 hours at hourly resolution. Models were trained individually for each of 52 stream gauges using non-public streamflow data from the 5-month monsoon seasons (June through October) in 2014 through 2019, and deployed during the 5-month monsoon season in 2020. We monitor several water level-based performance metrics -- the aggregate performance across the 52 watersheds are:

\begin{itemize}
\item Average lead time: 20.7 hours
\item R\textsuperscript{2} over water level: 0.99
\item Mean absolute error: 0.067 meters
\item Mean squared error: 0.011 meters\textsuperscript{2}
\end{itemize}

\subsection{Research model}
To improve on the deployed models, and to allow for launching at more locations during the upcoming 2021 monsoon season in India and Bangladesh, we tested the LSTM-based rainfall-runoff modeling strategy developed and benchmarked by \cite{kratzert2019towards}. LSTMs generally require large training data sets spanning multiple catchments \cite{gauch2020proper}, which exacerbates the challenge of using water level data with its strong dependence on catchment-specific bathymetry. Using high-quality precipitation and streamflow data from the US, we adapted the LSTM benchmarking experiments by \cite{kratzert2019towards} to train with water level data (see Appendix \ref{appendix:stage} for full details on the experimental setup). We achieved a median accuracy over 499 test basins of $r^2=0.71$ compared to $r^2=0.73$ with discharge. This is a small loss in accuracy when using water level instead of discharge, especially when compared with similar losses that occur when calibrating traditional rainfall-runoff models using water level data (e.g., \cite{jian2017towards}). 

\begin{figure}[ht]
\includegraphics[width=\textwidth]{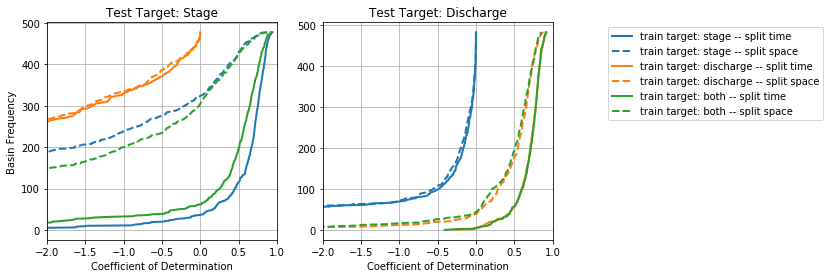} 
\centering
\caption{Cumulative density functions (CDFs) of model skill $(r^2)$ at predicting water level and discharge over 10-year test periods in 499 basins in the United States using models trained on water level and/or discharge and cross validated in time (i.e., at the same basins where they were trained) or space (i.e., at different basins than where they were trained).}
\label{fig:stage_cdfs}
\end{figure}

LSTMs trained on discharge data generalize relatively well to new catchments where streamflow data is not available for training \cite{kratzert2019toward} (most catchments in the world are ungauged). However, because the precipitation-stage relationship is heavily controlled by local bathymetry, models trained on stage data do not generalize in the same way as models trained on discharge. This is shown in Figure \ref{fig:stage_cdfs}, which plots cumulative density functions (CDFs) of results from applying LSTMs to simulate time series of daily streamflow (water level and discharge) from time series of daily precipitation at 499 US catchments. These CDFs compare coefficients of determination between 10 years of simulated vs. observed streamflow in each catchment when trained on water level data, discharge data, and both together using a multi-output head and a multi-target loss function. Models were tested in two types of prediction situations: (i) out-of-sample in time, and (ii) out-of-sample in space - the latter by using k-fold cross validation $(k=8)$ across 499 US catchments. Notice that the models trained on water level in gauged catchments (i.e., out-of-sample-in-time) are generally better than the ungauged (out-of-sample-in-space) discharge models, but that the ungauged (out-of-sample-in-space) discharge models are substantially better than the ungauged water level models.

% While LSTMs are almost as effective at simulating stage (as compared to simulating discharge) across a large number of diverse watersheds (a median $r^2$ of 0.71 vs. 0.73), models trained on stage data do not extrapolate well between catchments due to local differences in bathymetry. Discharge models lose some accuracy when predicting out-of-sample in space (see \cite{kratzert2019toward} for a full benchmark of this situation). Similarly, models trained on stage data do not predict discharge well, and vice-versa. 

These results show that to provide best-possible predictions everywhere, it is necessary to train models that predict both water level and discharge: discharge can be estimated with higher accuracy than river stage in ungauged basins, however the multi-target head and loss function allow to simultaneously train on water level data, and this provides benefit in catchments that have only water level data. Ideally, any inundation mapping strategy that relied on this type of rainfall-runoff model would be able to utilize both/either stage and discharge estimates.

\section{Morphological models}

%The above approach is extremely scalable and can be applied anywhere in the world with almost no effort, yet it has disadvantages in terms of accuracy. The pure data-driven approach, without the need for elevation maps, means that we can never properly forecast aspects of flooding that do not directly follow from things we've already seen. This is most obvious in unprecedented events, but is relevant to more frequent events as well - the approach above could not, for example, take into account recent changes to the riverbed morphology, new embankments or other infrastructure built since the last flood season.

Inundation models simulate the movement of water across the floodplain to produce spatially accurate forecasts of flood extent. They accept as an input hydrologic boundary conditions (expressed either as discharge or water level), and output a map of the floodplain inundation, including water depth at each pixel. 

The classic approach to implementing such models is using finite-element solutions to the St. Venant (or Shallow Water) equations. Such models are used almost universally in both academia and operational settings. However, there are many challenges in increasing their scale and reach. For a deeper discussion on these challenges, and efforts to address them within the scope of finite-element solutions, see \cite{ben2019inundation}.

In this section we present an approach to inundation modeling that improves on classic hydrodynamic modeling in many respects, including scalability, computational efficiency, and accuracy (at least in data-scarce settings). These models require an up-to-date, high resolution elevation map of the relevant floodplain (around one meter resolution, depending on the characteristics of the basin). They do not, however, require knowing riverbed bathymetry. %This is the main distinction in requirements between this methodology and the pure-ML model. % - information that is unavailable for the vast majority of rivers in the world, but of critical importance to classic hydrodynamic models.

In this approach, which we call \emph{the morphological model}, we break the inundation modeling task into two separate components:

\begin{itemize}

\item Learn the river profile -- i.e., the water level at every point along a one-dimensional line representing a river -- as a function of water level (stage) measurement at some point along the river.

\item Given the river profile, estimate inundation depth across the full two-dimensional floodplain.

\end{itemize}

These steps are described below in reverse order.

\subsection{Expanding the river profile to the floodplain}
\label{subsec:expand_to_flood_plain}

Assuming we know the water level at each point along a one-dimension river, we'd like to extend the water across the floodplain to identify what areas are inundated, and at what depth. We do so with no learned parameters -- by simply applying a deterministic algorithm that follows from certain basic principles and simple heuristics. Classically, this is done using a finite-element solution to the two-dimensional St. Venant equations, but that type of solution is computationally expensive, and is sensitive to both numerical instabilities and minor errors in the elevation map, leading to significant manual labor in identifying and correcting such issues. 

To avoid these hurdles, we employ a simple heuristic approach. To understand this approach, we begin with an inundation map that has no flow (i.e. the surface of the water is flat). In this degenerate case, given the height of the water at any point within the map, a simple 2D flood fill algorithm would yield the inundation map across the entire flood plain.

Next, in the simple case where the river is a straight line but the surface of the water is no longer flat, we assume that most of the dynamics of the downstream flow are already captured in the river profile. We assume that the water flow direction is approximately in the direction of the river, and so no (or little) flow occurs in directions that are perpendicular to the river. Thus, given the river profile, we can calculate the height of the water surface at any point in the map by simply expanding the river profile in the direction perpendicular to the river. Subtracting this height from every point in the elevation map results in a new elevation map, or a new floodplain morphology. In this new morphology, the water's surface must be flat, and thus we are back in the degenerate case (see Figure \ref{fig:flatten} in Appendix \ref{appendix:morphological} for a visualization).

In reality, rivers rarely progress in perfectly straight lines. Yet this algorithm can be generalized to any river course. We find constant elevation lines in an inundation map matching the given elevation map. This is equivalent to mapping each pixel in a 2D floodplain to a pixel in the 1D river profile that will have a similar water level in any inundation map. To do this, We use a simple heuristic which assigns each floodplain pixel with the river pixel closest to it, and then apply smoothing to make sure that pixels that are close to each other in the floodplain are not associated with pixels that are very distant from each other in the river (see Figure \ref{fig:associate} in Appendix \ref{appendix:morphological}). We believe this heuristic can be improved, though even it performs relatively well. Once this mapping is defined, we continue similarly to the original (straight line) case with the associated subsets of pixels replacing the perpendicular lines. We subtract from each subset of pixels associated with the same river profile point the water level at that river profile point. We get a new morphology for the whole floodplain, one in which applying a simple two-dimensional flood-fill algorithm with water level zero achieves a good approximation of the true inundation map.

\subsection{Estimating the river profile}

Now that we know how to expand the river profile into the full floodplain, we return to estimating the river profile itself from a single gauge. Our goal is to learn a function from a single real variable (the gauge measurement) to a 1-dimensional line (the river profile). Alternatively, we can see this as estimating the water level at a point as a function of both the (estimated or measured) streamflow water level and location along the river. We can make several assumptions which greatly reduce the space of possible river profiles - we know the function is continuous in both gauge measurement and location, we know it is monotonously decreasing in location (since rivers, generally speaking, flow downwards rather than upwards), and we know the function is monotonously increasing in water level (since one can generally assume that an increase in water in one point along the river will not indicate a decrease in the amount of water at another point upstream or downstream from it). These assumptions also allow us to transfer information about the river profile at one point to other points.

We learn the function between the gauge measurement and the river profile using historical inundation maps. We can score a function on a given historical flood event by expanding the river profile at the relevant gauge measurement to the floodplain as described in Section \ref{subsec:expand_to_flood_plain} and comparing the resulting flood extents. We search for the function that optimizes this score over a catalogue of historical inundation maps which were derived from SAR satellite imagery. We use a local search approach to find the optimal function, raising or lowering the river profile at a specific gauge measurement depending on whether it produces overflooding or underflooding, while maintaining the monotonicity and continuity constraints described above. See Figure \ref{fig:learned} for a visualization example of this type of learned river profiles.

\subsection{Overview and results}

Combining these two steps we find that based only on an elevation map, past gauge measurements, and past inundation maps, we can (a) learn a function from the gauge measurement to the full river profile, (b) edit the morphology of the floodplain based on the river profile we have deduced, and (c) calculate the inundation map (including inundation depths) extremely efficiently using a simple flood-fill algorithm that is applied to the new synthetic morphology.

To evaluate the performance of this methodology relative to classic finite-element solution models, we trained and evaluated the two models across 11 different regions in the Ganges and Brahmaputra basins in India. We evaluated the models across the metrics in Table \ref{tab:innundation_benchmarks}. As seen in Table \ref{inundation_comparison}, the morphological model outperforms the classic hydrodynamic model across all metrics with the exception of recall, in which the two are statistically indistinguishable. 

\begin{table}[t]
\caption{Inundation benchmarking results.}\label{inundation_comparison}
\begin{tabular}{ | p{3.1cm }|| p{1.7cm }| p{2.1cm }|p{2.0cm}|p{2.9cm}|  }
 \hline
 Model type & Avg. Recall & Avg. Precision & Avg. Manual Work Hours$^a$ & CPU Costs \\
 \hline
 \hline
 Hydrodynamic model & 71.4\%  & 72.7\% & 30 & 4-131 CPU \emph{years} \\
 \hline
 Morphological model & 71.3\% & 75.8\% & 4 & 100-1200 CPU \emph{hours} \\
 \hline
\end{tabular}
{$^a$\textit{This parameter was roughly estimated by the engineers working on both models, and was not accurately measured.}}
\label{tab:innundation_benchmarks}
\end{table}

We have launched this morphological model to cover 38,000 square kilometers and 44 million people. In our real-time operational systems, spanning 429 flood events in the 2020 Monsoon season we achieve the following metrics at a 64 meter resolution:

\begin{itemize}

\item Precision: 76.2\%

\item Recall: 77.6\%

\end{itemize}

Note that the methodology described in this section includes an assumption that there are no confluences or bifurcations in the river. This assumption is not strictly required and the methodology can be generalized to more complex river structures, however this is outside the scope of the current paper.

\section{Conclusions and impact}

The methodologies described in this paper reduce the manual labor required to launch operational flood forecasting at new sites over large areas. They reduce reliance on difficult-to-attain and costly-to-measure data (e.g., discharge and riverbed bathymetry), and they reduce computation costs. These approaches do so largely while improving or at least not significantly harming accuracy relative to standard approaches. As such, these techniques offer an important step toward scaling flood warning systems that have the potential to cover billions of people. 

In addition to developing and benchmarking these algorithms, we have also deployed them in dozens of real-time operational flood warning systems - proving they work well in a real-world operational setting. Over the past 5 months we have shown this work can achieve real impact by sending more than 36 million warnings and alerts to individuals at risk from floods, as well as the relevant authorities to help support relief and mitigation efforts. We continue to work towards scaling these systems up further.

\section{Acknowledgements}
The authors would like to thank the many contributors to the flood forecasting system described in this paper: Adi Gerzi Rosenthal, Asher Metzger, Chen Barshai, Dana Weitzner, Efrat Morin, Gregory Begelman, Hila Noga, Ira Shavitt, Moriah Royz, Niv Giladi, Ofir Reich, Tal Shechter, Vova Anisimov, Yotam Gigi, Zach Moshe,  Zvika Ben-Haim.

\pagebreak

\bibliography{neurips_2020}
\bibliographystyle{unsrt}

\pagebreak

\begin{appendices}
\section{Water level rainfall-runoff model experiments}
\label{appendix:stage}

Experiments from Section \ref{section::stage} used the benchmarking protocol developed by \cite{kratzert2019towards}. Specifically, we used the Catchment Attributes and Meteorological Large Sample (CAMELS) data set curated by the US National Center for Atmospheric Research \cite{Addor2017camels, Neuman2015camels}. This data set consists of 671 catchments in CONUS ranging in size from 4 $km^2$ to 25,000 $km^2$ that have largely natural flows and long streamflow gauge records (1980-2010). We used only 499 of the CAMELS catchments - specifically, the basins with sub-daily gauge data available from the US Geological Survey.  

Inputs into the LSTM models included time series of meteorological variables from NASA's North American Land Data Assimilation System (precipitation, long- and short-wave radiation, wind speed, potential energy and potential evaporation, specific humidity, air temperature, and near-surface atmospheric pressure). Inputs also included static catchment attributes related to soils, climate, vegetation, topography, and geology \cite{Addor2017camels}. A full list of catchment attributes used as model inputs can be found in Table 1 by \cite{kratzert2019towards}.

Daily streamflow records (water level and/or discharge) were used as training targets with a normalized squared-error loss function that does not depend on basin-specific mean:
\begin{equation}
    \textup{NSE*} = \frac{1}{B} \sum_{b=1}^{B} \sum_{n=1}^{N} \frac{(\widehat{y}_n - y_n)^2}{(s(b) + \epsilon)^2},
\end{equation}
This loss function means that larger basins, which tend to have more discharge, are not over-represented in the effect on weight updates during training. The multi-target loss function was an equally-weighted average of this loss function over both variables (stage and discharge).

The training period was from October 1, 1999 to September 30, 2008 and the testing period was from October 1, 1989 through September 1, 1999. The LSTM used sequence-to-one prediction with a 365-day lookback, which increases minibatch diversity and the number of weight updates per epoch relative to sequence-to-sequence training.

\section{Morphological model visualizations}
\label{appendix:morphological}

This appendix provides several visuals that help make the morphological model more intuitive.

\begin{figure}[ht]
\includegraphics[width=\textwidth]{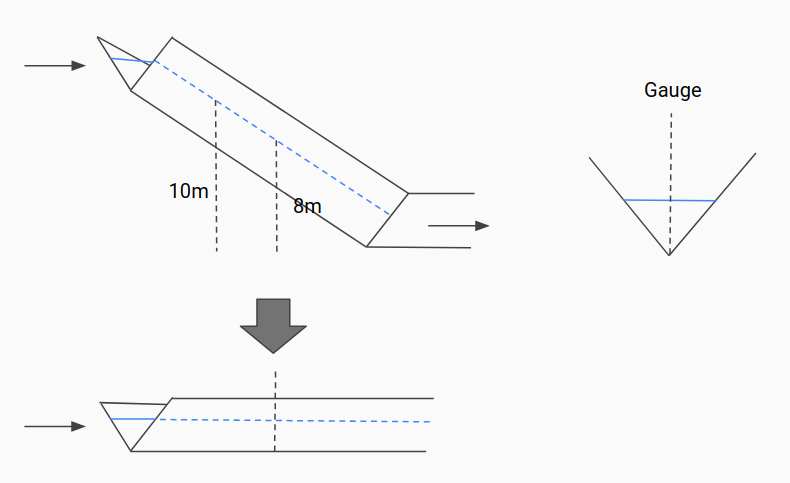} 
\centering
\caption{A simple example of a river being flattened so that the river profile is completely level.}
\label{fig:flatten}
\end{figure}

Figure~\ref{fig:flatten} illustrates a simple example of a river being flattened. The river, represented in three dimensions at the top of the image, is flattened by subtracting the water level at each point from the elevation map at that point. Regardless of the original structure, the resulting river will always be completely flat.

\begin{figure}[ht]
\includegraphics[width=\textwidth]{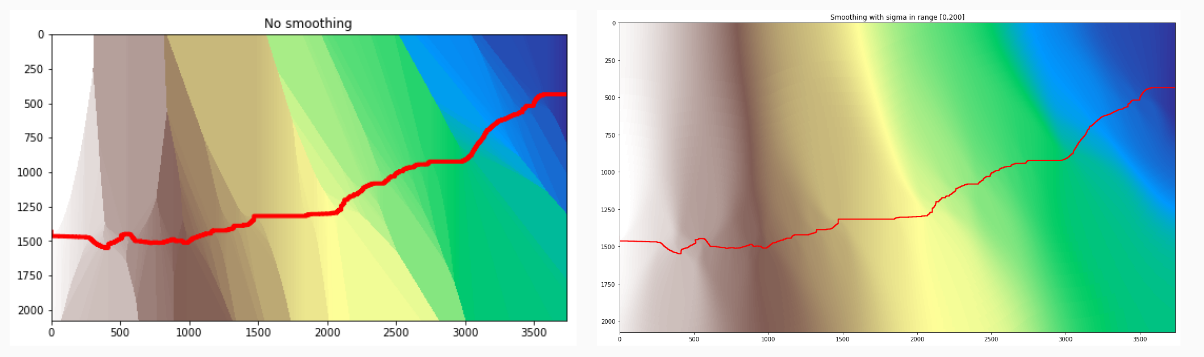} 
\centering
\caption{Pixels in the floodplain being associated with points along the river profile, on the left without smoothing and on the right with smoothing.}
\label{fig:associate}
\end{figure}

Figure~\ref{fig:associate} shows how we associate pixels of the floodplain to points along the river profile, expressed as a color from white to blue. On the left we see the result if we simply associate each floodplain pixel with the closest river pixel, which leads to significant discontinuities which are not physically sensible. On the right we apply smoothing to the association, leading to a reasonable approximation of the accurate association.

\begin{figure}[ht]
\includegraphics[width=\textwidth]{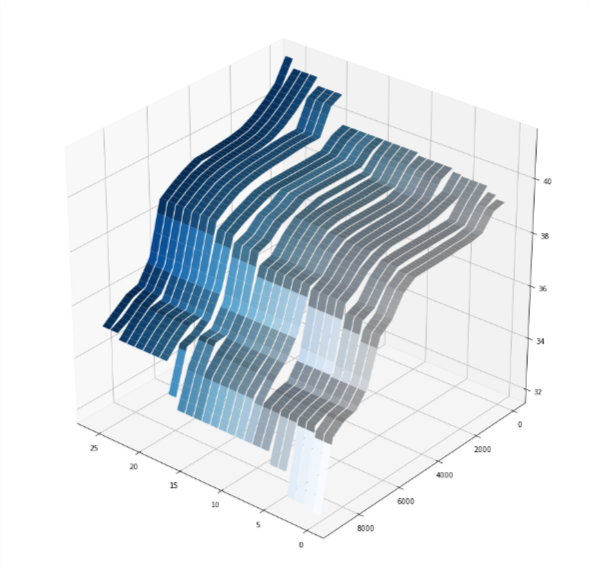} 
\centering
\caption{Learned river profiles, represented as water level as a function of both the location along the river and the gauge measurement.
The x axis refers to 27 increasing gauge measurements.
The y axis represents the distance in meters from a reference point in the river in steps of 16 meters along the river curve.
The z axis represents the water level in meters at the specific point in the river and gauge measurement.}
\label{fig:learned}
\end{figure}

Figure~\ref{fig:learned} shows an example of learned river profiles in the Brahmaputra river in India, showing the water level at each point (up to 50 meters) as a function of both the location of the point along the river (up to 10,000 meters) and the gauge measurement at the time (up to 30 meters).

\end{appendices}
\end{document}